\documentclass[12pt]{article}
\usepackage{amsmath}
\usepackage{graphicx}
\usepackage{axodraw}
\begin{document}
\baselineskip 0.6cm

\def\simgt{\mathrel{\lower2.5pt\vbox{\lineskip=0pt\baselineskip=0pt
           \hbox{$>$}\hbox{$\sim$}}}}
\def\simlt{\mathrel{\lower2.5pt\vbox{\lineskip=0pt\baselineskip=0pt
           \hbox{$<$}\hbox{$\sim$}}}}

\begin{titlepage}

\begin{flushright}
UCB-PTH-11/09 \\
\end{flushright}

\vskip 1.5cm

\begin{center}

{\Large \bf 
Spread Supersymmetry
}

\vskip 0.8cm

{\large Lawrence J. Hall and Yasunori Nomura}

\vskip 0.4cm

{\it Berkeley Center for Theoretical Physics, Department of Physics, \\
     and Theoretical Physics Group, Lawrence Berkeley National Laboratory, \\
     University of California, Berkeley, CA 94720, USA} \\

\abstract{In the multiverse the scale of supersymmetry breaking, 
 $\tilde{m} = F_X/M_*$, may scan and environmental constraints on the 
 dark matter density may exclude a large range of $\tilde{m}$ from the 
 reheating temperature after inflation down to values that yield a lightest 
 supersymmetric particle (LSP) mass of order a TeV.  After selection 
 effects, for example from the cosmological constant, the distribution 
 for $\tilde{m}$ in the region that gives a TeV LSP may prefer larger 
 values.  A single environmental constraint from dark matter can then 
 lead to multi-component dark matter, including both axions and the LSP, 
 giving a TeV-scale LSP somewhat lighter than the corresponding value 
 for single-component LSP dark matter.

 If supersymmetry breaking is mediated to the Standard Model sector at 
 order $X^\dagger X$ and higher, only squarks, sleptons and one Higgs 
 doublet acquire masses of order $\tilde{m}$.  The gravitino mass is 
 lighter by a factor of $M_*/M_{\rm Pl}$ and the gaugino masses are 
 suppressed by a further loop factor.  This Spread Supersymmetry spectrum 
 has two versions, one with Higgsino masses arising from supergravity 
 effects of order the gravitino mass giving a wino LSP, and another 
 with the Higgsino masses generated radiatively from gaugino masses 
 giving a Higgsino LSP.  The environmental restriction on dark matter 
 fixes the LSP mass to the TeV domain, so that the squark and slepton 
 masses are order $10^3~{\rm TeV}$ and $10^6~{\rm TeV}$ in these two 
 schemes.  We study the spectrum, dark matter and collider signals of 
 these two versions of Spread Supersymmetry.  The Higgs boson is Standard 
 Model-like and predicted to lie in the range $110~\mbox{--}~145~{\rm GeV}$; 
 monochromatic photons in cosmic rays arise from dark matter annihilations 
 in the halo; exotic short charged tracks occur at the LHC, at least 
 for the wino LSP; and there are the eventual possibilities of direct 
 detection of dark matter and detailed exploration of the TeV-scale 
 states at a future linear collider.  Gauge coupling unification is 
 at least as precise as in minimal supersymmetric theories.

 If supersymmetry breaking is also mediated at order $X$, a much less 
 hierarchical spectrum results.  The spectrum in this case is similar 
 to that of the Minimal Supersymmetric Standard Model, but with the 
 superpartner masses $1~\mbox{--}~2$ orders of magnitude larger than 
 those expected in natural theories.}

\end{center}
\end{titlepage}

\section{Introduction}
\label{sec:intro}

The physical origin of the weak scale is currently being probed at the 
Large Hadron Collider (LHC).  In the Standard Model (SM), the weak scale 
is unnatural; in particular, the Higgs mass parameter must be fine-tuned 
by many orders of magnitude, leading to the almost universal belief that 
some new physics must be rigidly connected to the weak scale, $v \sim 
200~{\rm GeV}$.  Supersymmetry, a rather unique and elegant extension 
of spacetime symmetry, has emerged as the leading candidate for this new 
physics, with the scale of weak interactions linked directly to the scale 
of supersymmetric particles, $\tilde{m}$.

However, the multiverse, suggested by the plethora of string theory 
vacua~\cite{Bousso:2000xa}, casts doubt on this picture.  The overall 
magnitude of the weak scale may be selected by anthropic requirements 
for an observable universe.  Denying a symmetry explanation for the 
weak scale is controversial, but the exploration of an anthropic 
weak scale is well-motivated.  While the physical effects that might 
restrict the size of the weak scale are not fully clear, we know that 
changing $v$ by a factor of a few will lead to drastic changes of 
the universe~\cite{Agrawal:1997gf}.  Indeed, a similar argument 
applied to the cosmological constant has led to the successful 
understanding of the order of magnitude of the observed dark 
energy~\cite{Weinberg:1987dv,Martel:1997vi}---something that has 
not been achieved using a symmetry argument.

Even if the weak scale is anthropically determined, the underlying theory 
may still be supersymmetric---after all, supersymmetry is necessary 
for the consistency of string theory.  We are therefore faced with the 
interesting possibility that $\tilde{m}$ and $v$ are not rigidly coupled 
to each other, but rather are decoupled~\cite{ArkaniHamed:2004fb}.  A 
fundamental Higgs boson with a mass very far below $\tilde{m}$ would 
imply many orders of magnitude of fine-tuning which would be strong 
evidence for environmental selection of the weak scale.  But with 
supersymmetry breaking out of reach of high energy colliders, what 
observations could confirm this picture?

Two schemes that allow precision tests of a finely tuned weak scale far 
below $\tilde{m}$ are Split Supersymmetry~\cite{ArkaniHamed:2004fb} and 
High Scale Supersymmetry~\cite{Hall:2009nd}.  In Split Supersymmetry 
the scalar superpartners have masses near $\tilde{m}$ while the fermionic 
superpartners are taken to have masses of order the TeV scale to account 
for the observed dark matter of the universe.  Measurements of the gluino 
lifetime, the gauge Yukawa couplings of the Higgsinos, and the Higgs 
boson mass would all be correlated with $\tilde{m}$.  Furthermore, such 
a correlation would imply that the Higgs is elementary up to $\tilde{m}$, 
so that the demonstration of fine-tuning would be convincing.  Split 
Supersymmetry yields gauge coupling unification with precision comparable
to the Minimal Supersymmetric Standard Model (MSSM).

In High Scale Supersymmetry all superpartners have masses of order 
$\tilde{m}$ and are inaccessible to colliders.  Nevertheless, for 
$\tilde{m} \simgt 10^{11}~{\rm GeV}$ the Higgs boson mass is predicted 
to be in the range of $(128~\mbox{--}~141)~{\rm GeV}$, depending on the 
composition of the Higgs boson.  Moreover, many theories lead to values 
of the Higgs mass at the edge of this range, and the UV uncertainties 
from the superpartner spectrum are extremely small, less than 
$0.5~{\rm GeV}$.  In High Scale Supersymmetry it is the value of 
the Higgs boson mass itself that provides evidence of many orders 
of magnitude of fine-tuning.  High Scale Supersymmetry can be made 
complete with axion dark matter; if $f_a$ is above $10^{12}~{\rm GeV}$, 
as typically expected, the axion misalignment angle is environmentally 
selected to be small~\cite{Linde:1987bx,Tegmark:2005dy}.  Gauge 
coupling unification, while less precise than in the MSSM, is 
nevertheless still significant.

In this paper we pursue two related studies.  First we argue that in 
theories with $\tilde{m}$ varying in the multiverse, an environmental 
requirement from Large Scale Structure forbids a very large window 
of $\tilde{m}$.  This corresponds to the range in which the Lightest 
Observable-sector Supersymmetric Particle (LOSP) has a mass between 
order TeV and the reheating temperature of the universe after inflation 
$T_R$.  For large values of $T_R$, this forbidden region divides 
theories into two very separate classes:\ those with (some) superpartners 
at the TeV scale and those without.  Note that even in the former 
case, these superpartner masses are not directly linked to the scale 
of electroweak symmetry breaking and their existence is not needed 
to comprise dark matter.  The Lightest Supersymmetric Particle 
(LSP), which may or may not be the LOSP, is cosmologically stable 
and contributes to dark matter; but typically we expect multi-component 
dark matter with a significant axion contribution.  One member of this 
class has the entire MSSM in the TeV domain.  This Environmental MSSM 
will typically have several orders of magnitude of fine-tuning in weak 
symmetry breaking, and likely accounts for only a fraction of dark matter.

Secondly, within the class of theories where the environmental constraint 
forces at least some superpartners to be at the TeV scale, we perform 
a top-down analysis to arrive at models with a very simple theoretical 
structure:\ Spread Supersymmetry.  A key feature of this scheme is that 
the underlying structure of the theory forces a large spread in the 
superpartner spectrum so that, even though the LOSP ($=$ LSP in this 
case) is in the TeV domain, $\tilde{m}$ and most of the superpartner 
spectrum are several orders of magnitude larger than the TeV scale. 
Measurements on the few superpartners that have TeV scale masses 
correlate with the value of the Higgs boson mass, allowing $\tilde{m}$ 
to be inferred and yielding evidence for a very high degree of fine-tuning.

The scheme of Spread Supersymmetry postulates that a chiral supermultiplet 
$X$ responsible for supersymmetry breaking is charged under some symmetry, 
so that supersymmetry breaking is transferred to the MSSM sector via 
operators involving $X^\dagger X$, but not via operators linear in 
$X$, which would generate the gaugino masses, the supersymmetric Higgs 
mass ($\mu$ term), and the scalar trilinears ($A$ terms).  The leading 
supersymmetry breaking effects, therefore, arise from
\begin{equation}
  {\cal L}_{\rm SB} \,\sim\, \frac{1}{M_*^2}\, [X^\dagger X\, 
    (Q^\dagger Q + U^\dagger U + D^\dagger D + L^\dagger L + E^\dagger E 
    + H_u^\dagger H_u + H_d^\dagger H_d 
    + H_u H_d)]_{\theta^4},
\label{eq:susybr}
\end{equation}
where $\{Q,U,D,L,E\}$ and $\{H_u,H_d\}$ are the matter and Higgs 
superfields, and $M_*$ is the scale at which supersymmetry breaking 
is mediated to the MSSM sector.  The supersymmetry breaking masses 
generated by these operators are of order $\tilde{m} \equiv F_X/M_*$. 
Note that each operator has an unknown coefficient of order unity 
that is not displayed.  The coefficient of the last term could be 
suppressed due to an approximate Peccei-Quinn (PQ) symmetry.

The gauginos and Higgsinos acquire masses from higher order effects 
in $F_X/M_*^2$ or through $R$ symmetry breaking necessary to suppress 
the cosmological constant in supergravity.%
\footnote{We assume that the supersymmetry preserving vacuum expectation 
 value $\langle X \rangle$ is sufficiently small that contributions to 
 gaugino and Higgsino masses from operators involving $X^\dagger X$ are 
 negligible.}
This yields a spectrum for superpartners that spans some range.  Since 
$\tilde{m}$ scans in the multiverse, the forbidden window in the LOSP 
mass from the TeV scale to $T_R$ will force either the LOSP to be 
heavier than $T_R$ or to be in the TeV domain.  The former simply 
gives a perturbation of High Scale Supersymmetry, but the latter 
leads to Spread Supersymmetry.

\begin{figure}[t]
\begin{center}
\begin{picture}(415,270)(0,-20)
  \LongArrow(0,-10)(0,230) \Text(0,235)[b]{mass [TeV]}
  \Line(-3,0)(3,0) \Line(-3,30)(3,30) \Text(-8,30)[r]{\small $1$}
  \Line(-3,60)(3,60) \Line(-3,90)(3,90)
  \Line(-3,120)(3,120) \Text(-6,122)[r]{\small $10^3$}
  \Line(-3,150)(3,150) \Line(-3,180)(3,180)
  \Line(-3,210)(3,210) \Text(-6,212)[r]{\small $10^6$}
  \Line(20,21)(60,21) \Text(70,23)[l]{$\tilde{h}$}
  \Line(20,88)(60,88) \Text(70,86)[l]{$\tilde{W}$}
  \Line(20,95)(60,95) \Text(70,98)[l]{$\tilde{B}$}
  \Line(20,105)(60,105) \Text(72,110)[l]{$\tilde{g}$}
  \Line(20,165)(60,165) \Text(70,167)[l]{$\tilde{G}$}
  \Line(20,191)(60,191) \Line(20,193)(60,193) \Line(20,195)(60,195) 
  \Line(20,197)(60,197) \Line(20,199)(60,199)
  \Text(70,197)[l]{$\tilde{q}, \tilde{\ell}, H^{0,\pm}, A$}
  \DashLine(0,31.2)(125,31.2){3}
  \Text(133,31.2)[l]{$\Omega_{\tilde{h}} = \Omega_{\rm DM}$}
  \LongArrow(250,-10)(250,230) \Text(250,235)[b]{mass [TeV]}
  \Line(247,0)(253,0) \Line(247,30)(253,30) \Text(242,30)[r]{\small $1$}
  \Line(247,60)(253,60) \Line(247,90)(253,90)
  \Line(247,120)(253,120) \Text(244,122)[r]{\small $10^3$}
  \Line(247,150)(253,150) \Line(247,180)(253,180)
  \Line(247,210)(253,210) \Text(244,212)[r]{\small $10^6$}
  \Line(270,25)(310,25) \Text(320,24)[l]{$\tilde{W}$}
  \Line(270,33)(310,33) \Text(320,36)[l]{$\tilde{B}$}
  \Line(270,52)(310,52) \Text(322,54)[l]{$\tilde{g}$}
  \Line(270,105)(310,105) \Line(270,110)(310,110)
  \Text(320,108)[l]{$\tilde{h}, \tilde{G}$}
  \Line(270,133)(310,133) \Line(270,135)(310,135) \Line(270,137)(310,137) 
  \Line(270,139)(310,139) \Line(270,141)(310,141)
  \Text(320,139)[l]{$\tilde{q}, \tilde{\ell}, H^{0,\pm}, A$}
  \DashLine(250,43.9)(375,43.9){3}
  \Text(383,43.9)[l]{$\Omega_{\tilde{W}} = \Omega_{\rm DM}$}
\end{picture}
\caption{Two versions of the Spread Supersymmetry spectrum, with the 
 Higgsino LSP (left) and the wino LSP (right).}
\label{fig:spectrum}
\end{center}
\end{figure}
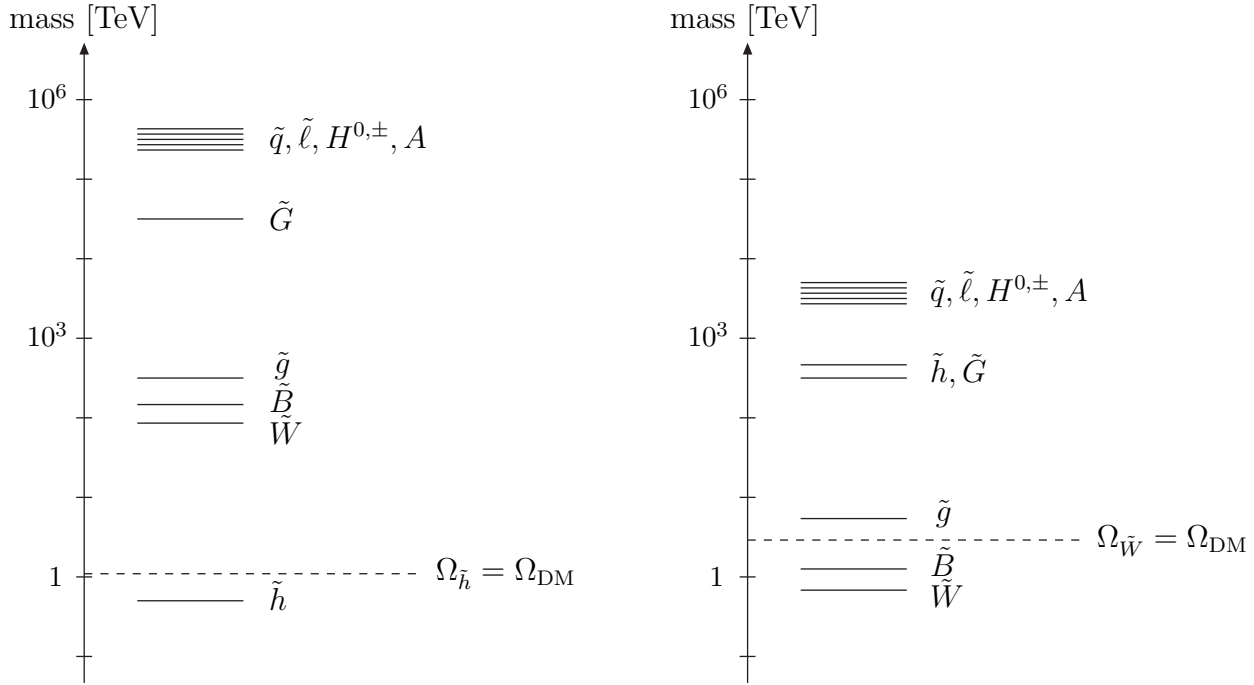
The spread spectrum we have in mind is illustrated in Fig.~\ref{fig:spectrum}. 
The gravitino mass $m_{3/2} = F_X/\sqrt{3}M_{\rm Pl} \equiv \epsilon_* 
\tilde{m}$ breaks $R$ symmetry and anomaly mediation leads to gaugino 
masses of order $m_{3/2}/16\pi^2$, so that the gauginos are lighter than 
the squarks and sleptons by a factor $\epsilon_*/16\pi^2$. ($\epsilon_* 
\equiv M_*/\sqrt{3}M_{\rm Pl}$ is typically smaller than $1$.)  The only 
remaining question is the mass of the Higgsinos, which is model dependent. 
Two versions of the spread spectrum are shown in Fig.~\ref{fig:spectrum}. 
In the left panel the Higgsino masses arise from a one-loop radiative 
correction form virtual gauginos and Higgs bosons.  In the right panel 
the Higgsino masses are of order the gravitino mass, which can arise from 
supergravity interactions that follow from having $H_u H_d$ in the 
K\"{a}hler potential~\cite{Giudice:1988yz} or that cause a readjustment 
of the vacuum~\cite{Hall:1983iz,Hempfling:1994ae}.  The overall spread 
of the superpartner spectrum is typically $3~\mbox{--}~6$ orders of 
magnitude.  The normalization of the spectrum is set by environmental 
selection which forces the LSP mass below a critical value at the TeV 
scale.  In both cases the spectra are sufficiently heavy to solve the 
supersymmetric flavor, $CP$ and cosmological gravitino problems.  The 
spectra shown are for $\epsilon_* \sim 10^{-2}$, corresponding to a high 
messenger scale of supersymmetry breaking of $M_* \sim 10^{16}~{\rm GeV}$. 
Since the normalization of the spectra is determined by the environmental 
selection of dark matter, as $M_*$ (and so $\epsilon_*$) is reduced the 
only effect is to raise the masses of the squarks, sleptons and heavy 
Higgs---the gravitino, Higgsino and gaugino masses are not affected. 
For definiteness we will consider $\epsilon_* \sim 10^{-2}$ in the rest 
of the paper.

In the next section, we explore in a fairly general setting how 
environmental selection can exclude a very large range of $\tilde{m}$. 
In section~\ref{sec:HiggsinoLSP}, we study Spread Supersymmetry with 
the LSP Higgsino spectrum of the left panel of Fig.~\ref{fig:spectrum}. 
We elucidate the theoretical structure of the model and study the 
phenomenology of the Higgsino states at colliders and for dark matter. 
In section~\ref{sec:WinoLSP} we introduce and study the model of 
Spread Supersymmetry with the LSP wino spectrum of the right panel 
of Fig.~\ref{fig:spectrum}.  For both theories we pay attention to 
the Higgs boson mass prediction, and we also point out that in both 
cases the precision of gauge coupling unification is comparable to 
that of the MSSM.  In section~\ref{sec:E-MSSM} we comment on the 
Environmental MSSM.  In section~\ref{sec:mu-scanning} we consider 
the case in which $\tilde{m}$ and $\mu$ scan independently.  Finally, 
we conclude in section~\ref{sec:concl}.

\section{The Forbidden Window in {\boldmath $\tilde{m}$}}
\label{sec:forb-window}

We consider any supersymmetric theory where the ratio of superpartner 
masses is fixed, but the overall scale of supersymmetry breaking, 
$\tilde{m} = F_X/M_*$, scans.  Furthermore, we assume that the LSP is 
cosmologically stable, and that the reheating temperature after inflation, 
$T_R$, does not scan.  In the multiverse, the value of $\tilde{m}$ is 
determined by the prior distribution and anthropic selection.  We assume 
that there is an anthropically allowed region for the amount of dark 
matter abundance; candidate boundaries for the region are discussed 
in Ref.~\cite{Tegmark:2005dy} and used to limit the axion component 
of dark matter.  This affects selection of $\tilde{m}$, since the LSP 
relic abundance depends on $\tilde{m}$.

\begin{figure}[t]
\begin{center}
\begin{picture}(290,225)(-10,-45)
  \GBox(0,25)(260,160){0.8}
  \SetColor{White}
  \Line(0,160)(260,160) \Line(0,25)(260,25)
  \Line(0,25)(0,160) \Line(260,25)(260,160)
  \SetColor{Black}
  \LongArrow(0,-15)(0,160) \Text(-4,157)[rb]{\large $\xi_{\rm LSP}$}
  \LongArrow(-15,0)(260,0) \Text(260,-7)[t]{\large $m_{\rm LOSP}$}
  \Line(20,10)(190,130) \CArc(192.8,126.4)(4.5,4,128)
  \Line(197.3,126.7)(209,-10)
  \Text(-3,25)[r]{$\xi_{\rm DM,c}$}
  \DashLine(41.25,25)(41.25,-20){2} \Text(41.25,-27)[t]{\small $\sim {\rm TeV}$}
  \DashLine(206,25)(206,-20){2} \Text(206,-27)[t]{\small $\approx T_R$}
  \LongArrow(123.625,-10)(41.25,-10) \LongArrow(123.625,-10)(206,-10)
  \Text(123.625,-15)[t]{The forbidden window}
\end{picture}
\caption{A large window between $\sim {\rm TeV}$ and $\approx T_R$ for 
 the LOSP mass, $m_{\rm LOSP}$, is forbidden since it overproduces dark 
 matter, beyond the bound in Eq.~(\ref{eq:xi-bound}).}
\label{fig:forbidden}
\end{center}
\end{figure}
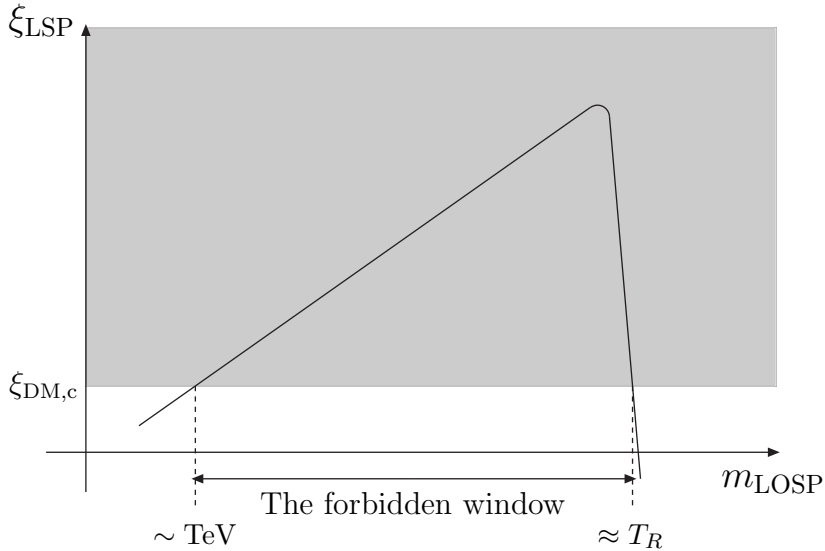

The LSP may or may not be the LOSP---if not (e.g.\ if it is the gravitino 
$\tilde{G}$) then its abundance is determined by late decay of the 
LOSP: $\rho_{\rm LSP} = (m_{\rm LSP}/m_{\rm LOSP}) \rho_{\rm LOSP}$. 
Here, $\rho_i$ is the energy density of species $i$, and we assume 
that $m_{\rm LSP}$ and $m_{\rm LOSP}$ are not many orders of magnitude 
different.  If the LOSP mass is below $T_R$, it is brought into thermal 
equilibrium and, as the temperature drops below its mass, freezes-out 
with an abundance $\xi_{\rm LOSP} = \rho_{\rm LOSP}/s$, where $s$ is the 
total entropy.  The resulting dark matter density is very high if the 
LOSP mass is well above a TeV, beyond the upper edge of the anthropically 
allowed region characterized by a critical dark matter abundance 
$\xi_{\rm DM,c}$.  This, therefore, leads to an environmental selection 
of $\tilde{m}$
\begin{equation}
  \xi_a + \xi_{\rm LSP}(\tilde{m}) \,<\, \xi_{\rm DM,c},
\label{eq:xi-bound}
\end{equation}
where $\xi_{\rm LSP} = (m_{\rm LSP}/m_{\rm LOSP}) \xi_{\rm LOSP}$, $\xi_a$ 
is the axion abundance that depends on parameters in the axion sector, and 
we assume that possible dependence of $\xi_{\rm DM,c}$ on $\tilde{m}$ is 
weak.  This single condition simultaneously selects for a low enough axion 
density~\cite{Linde:1987bx} and for a low LSP/LOSP mass.  Hence, there is 
an environmentally forbidden window for $\tilde{m}$ between the TeV scale 
and $T_R$, as illustrated in Fig.~\ref{fig:forbidden}---the LOSP must 
either be heavy enough so that it is not produced significantly after 
inflation, or it must be light enough to satisfy Eq.~(\ref{eq:xi-bound}).

The number of decades of this forbidden window increases with $T_R$ 
as $\log_{10}(T_R/{\rm TeV})$.  For values of $T_R \gg {\rm TeV}$, this 
window is very large and divides theories into two categories, those 
with no superpartners below $T_R$ and those with at least some superpartners 
in the TeV domain.  Many simple theories in the first category will yield 
High Scale Supersymmetry, with a Higgs mass prediction in the range of 
$(128~\mbox{--}~141)~{\rm GeV}$, depending on $\tan\beta$, providing 
$T_R$ is $10^{10}~{\rm GeV}$ or larger.  For theories in the second 
category, Large Scale Structure may not limit how light the LOSP can 
be, since dark matter may be fully accounted for by axions.  Since 
supersymmetry has not yet been discovered at colliders, this suggests 
that the multiverse distribution for $\tilde{m}$ in this region favors 
larger values, so that the LOSP mass is near the edge of the forbidden 
window.  This second category of theories we call TeV-LOSP Supersymmetry.

The probability distribution for observing a universe with 
supersymmetry breaking $\tilde{m}$ may then be written as $f(\tilde{m})\, 
\theta(\tilde{m})\, d\tilde{m}$, where $\theta$ is unity (vanishing) 
for values of $\tilde{m}$ outside (inside) the forbidden window, and 
$f$ contains effects from the a priori distribution of $\tilde{m}$ in 
the multiverse and environmental selection other than Eq.~(\ref{eq:xi-bound}).%
\footnote{For $\tilde{m} \gg v$ the fine-tuning necessary for electroweak 
 symmetry breaking contributes a factor $\tilde{m}^{-2}$ to $f(\tilde{m})$. 
 If this were the only effect, the a priori distribution must be peaked 
 more strongly than $\tilde{m}^2$ in order for $f(\tilde{m})$ to prefer 
 larger $\tilde{m}$ values.  It is, however, likely that other effects 
 also contribute~\cite{Hall:2007ja}.  For example, if the cosmological 
 constant scans and is selected by the environmental condition of 
 Ref.~\cite{Weinberg:1987dv}, it strongly favors a large dark matter 
 density---for single component LSP dark matter it provides a factor 
 $\tilde{m}^8$ to $f(\tilde{m})$---and it is likely to overwhelm both 
 the a priori distribution and the effect from Higgs fine-tuning, 
 yielding TeV-LOSP supersymmetry with dark matter close to the 
 environmental boundary.}
If $f$ increases with $\tilde{m}$ for LOSPs in the TeV region, does 
this mean that High Scale Supersymmetry is more likely than TeV-LOSP 
Supersymmetry?  No.  Even if $f$ increases very rapidly at low $\tilde{m}$, 
it is possible that $f$ transitions to a decreasing (or less rapidly 
growing) function of $\tilde{m}$ somewhere in the window because the 
forbidden region is very large.  In particular, since the two anthropically 
allowed regions are very distant from each other, the size of the 
anthropic factor in the two regions may differ by many orders of 
magnitude.  For example, if our universe has TeV scale superpartners, 
increasing $\tilde{m}$ by many orders of magnitude will lead to a large 
change in the size of the low energy gauge couplings.  Normalizing 
particle physics relative to the size of the QCD scale, this leads 
to a very large change in the strength of the gravitational interaction. 
The bottom line is that we do not know the relative probabilities of 
universes that are far apart in parameter space.

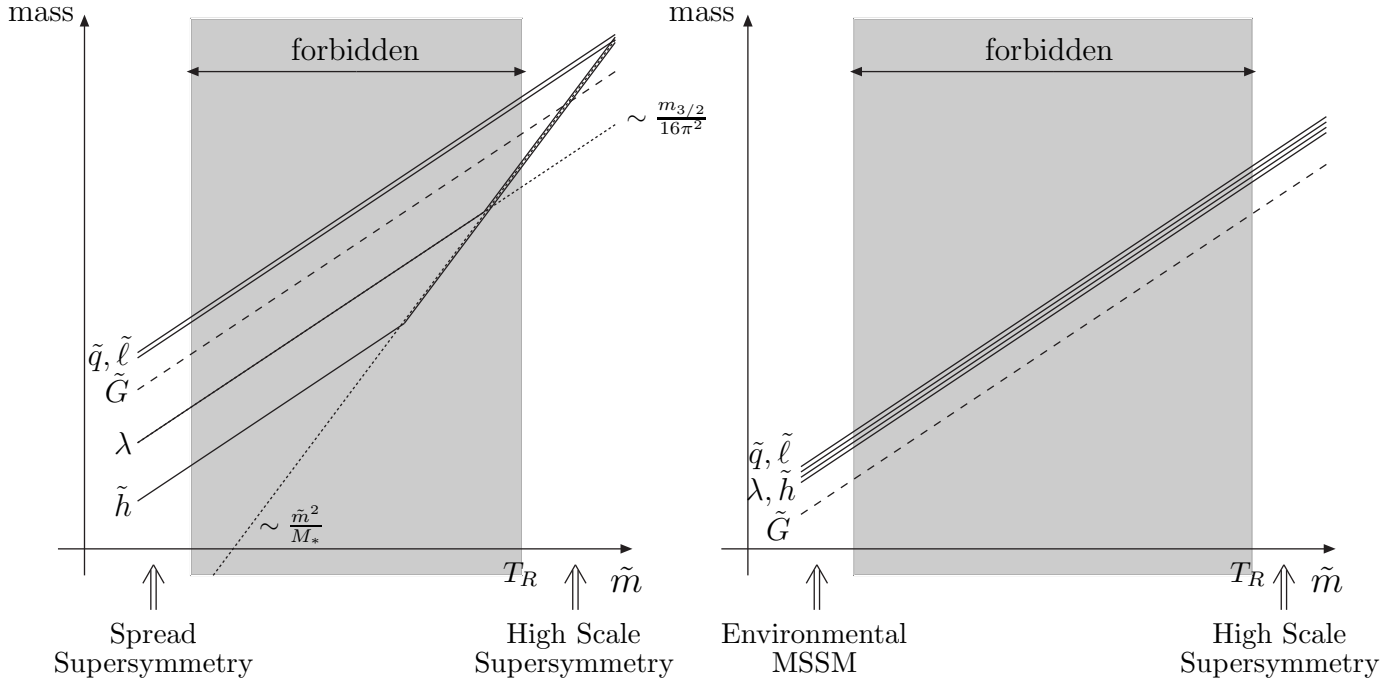
\begin{figure}[t]
\begin{center}
\begin{picture}(470,275)(-10,-60)
  \GBox(30,-10)(155,200){0.8}
  \SetColor{White}
  \Line(30,200)(155,200) \Line(30,-10)(155,-10)
  \Line(30,-10)(30,200) \Line(155,-10)(155,200)
  \SetColor{Black}
  \LongArrow(92.5,180)(30,180) \LongArrow(92.5,180)(155,180)
  \Text(92.5,185)[b]{forbidden}
  \Text(155,-5)[t]{\small $T_R$}
  \LongArrow(-10,-10)(-10,200) \Text(-14,200)[rb]{mass}
  \LongArrow(-20,0)(195,0) \Text(195,-7)[t]{\large $\tilde{m}$}
  \Line(10,72)(190,192) \Line(10,74)(190,194)
  \Text(7,74)[r]{$\tilde{q}, \tilde{\ell}$}
  \DashLine(10,60)(190,180){3} \Text(7,61)[r]{$\tilde{G}$}
  \DashLine(38.5,-10)(190,192){1}
  \Text(56,9)[l]{\footnotesize $\sim \frac{\tilde{m}^2}{M_*}$}
  \DashLine(10,40)(190,160){1}
  \Text(195,163)[l]{\footnotesize $\sim \frac{m_{3/2}}{16\pi^2}$}
  \Line(10,40)(140.5,127) \Line(140.5,127)(190,193) \Text(8,40)[r]{$\lambda$}
  \Line(10,18)(110.5,85) \Line(110.5,85)(190,191) \Text(8,18)[r]{$\tilde{h}$}
  \Line(15,-23)(15,-8)\Line(17,-23)(17,-8)
  \Line(16,-6)(12,-13) \Line(16,-6)(20,-13)
  \Text(16,-28)[t]{\small Spread} \Text(16,-39)[t]{\small Supersymmetry}
  \Line(174,-23)(174,-8)\Line(176,-23)(176,-8)
  \Line(175,-6)(171,-13) \Line(175,-6)(179,-13)
  \Text(175,-28)[t]{\small High Scale} \Text(175,-39)[t]{\small Supersymmetry}
  \GBox(280,-10)(430,200){0.8}
  \SetColor{White}
  \Line(280,200)(430,200) \Line(280,-10)(430,-10)
  \Line(280,-10)(280,200) \Line(430,-10)(430,200)
  \SetColor{Black}
  \LongArrow(355,180)(280,180) \LongArrow(355,180)(430,180)
  \Text(355,185)[b]{forbidden}
  \Text(430,-5)[t]{\small $T_R$}
  \LongArrow(240,-10)(240,200) \Text(236,200)[rb]{mass}
  \LongArrow(230,0)(460,0) \Text(460,-7)[t]{\large $\tilde{m}$}
  \Line(260,25)(458,157) \Line(260,27)(458,159)
  \Line(260,29)(458,161) \Line(260,31)(458,163)
  \Text(257,37)[r]{$\tilde{q}, \tilde{\ell}$}
  \Text(259,23)[r]{$\lambda, \tilde{h}$}
  \DashLine(260,13)(458,145){3} \Text(257,9)[r]{$\tilde{G}$}
  \Line(265,-23)(265,-8)\Line(267,-23)(267,-8)
  \Line(266,-6)(262,-13) \Line(266,-6)(270,-13)
  \Text(266,-28)[t]{\small Environmental} \Text(266,-39)[t]{\small MSSM}
  \Line(441,-23)(441,-8)\Line(443,-23)(443,-8)
  \Line(442,-6)(438,-13) \Line(442,-6)(446,-13)
  \Text(442,-28)[t]{\small High Scale} \Text(442,-39)[t]{\small Supersymmetry}
\end{picture}
\caption{The masses of superpartners as a function of $\tilde{m}$, both 
 on logarithmic scales, for the case where supersymmetry breaking is 
 transmitted to the superpartners at order $X^\dagger X$ (left) and at 
 order $X$ (right).  The forbidden window of Section~\ref{sec:forb-window} 
 is indicated by the shaded areas.}
\label{fig:tilde-m}
\end{center}
\end{figure}
In Fig.~\ref{fig:tilde-m} we illustrate how the superpartner spectrum 
depends on $\tilde{m}$ for a spread spectrum (left panel), where 
supersymmetry breaking is transmitted to the superpartners at order 
$X^\dagger X$, and for a normal spectrum (right panel), where the 
transmission of supersymmetry breaking is linear in $X$.  In the left 
panel we have chosen to show the spectrum corresponding to Spread 
Supersymmetry with a Higgsino LSP.   For $\tilde{m}$ well below the 
cutoff scale $M_*$, the spectra are simply proportional to $\tilde{m}$ 
but, as $\tilde{m}$ approaches $M_*$, higher dimension operators, 
giving the masses contributions of order $F_X^2/M_*^3 = \tilde{m}^2/M_*$, 
decrease the splitting of the spread spectrum and eventually the 
effective theory below $M_*$ ceases to be supersymmetric.

The forbidden window is shown on each panel of Fig.~\ref{fig:tilde-m} 
and corresponds to LOSP masses in the TeV to $T_R$ range.  Theories 
with $\tilde{m}$ above the forbidden window are labeled as High Scale 
Supersymmetry in both panels.  Providing $T_R$ is large, the splitting 
of the superpartner spectrum provides only small corrections to the 
Higgs boson mass prediction.  Theories with $\tilde{m}$ below the 
forbidden window are labeled as Spread Supersymmetry in the left panel 
and Environmental MSSM in the right panel; they have very different 
experimental consequences.  Spread Supersymmetry has the advantage 
that the squarks and leptons are sufficiently heavy to solve both the 
supersymmetric flavor and $CP$ problems, and the gravitino is sufficiently 
heavy that there is no cosmological gravitino problem.  Both these 
theories preserve successful gauge coupling unification of the MSSM. 
We devote the next two sections to the Higgsino LSP and Wino LSP 
versions of Spread Supersymmetry, and we comment on Environmental 
MSSM in Section~\ref{sec:E-MSSM}.

\section{Spread Supersymmetry with Higgsino LSP}
\label{sec:HiggsinoLSP}

\subsection{Spectrum}
\label{subsec:spectrum}

In any scheme with supersymmetry breaking arising at quadratic order 
in $X$, as in Eq.~(\ref{eq:susybr}), squarks, sleptons and the heavy 
Higgs doublet, $H$, have masses of order $\tilde{m}$.  The gauginos 
obtain masses of order $\epsilon_* \tilde{m}/16\pi^2$ from anomaly 
mediation~\cite{Randall:1998uk,Giudice:1998xp}; more specifically, 
$M_i =  b_i g_i^2 \, m_{3/2} / 16 \pi^2$, where $(b_1, b_2, b_3) = 
(33/5, 1 ,-3)$ are the beta-function coefficients, and we have taken 
the phase convention that $M_{1,2}$ are real and positive.  The lightest 
gaugino, therefore, is the wino.

The crucial question is how the Higgsino mass arises.  In this section 
we assume that the supersymmetric term $H_u H_d$ is absent both in the 
K\"{a}hler potential and in the superpotential.  Since PQ symmetry is 
broken by the last term of Eq.~(\ref{eq:susybr}) (and that $R$ symmetry 
is broken by the gaugino masses), the Higgsino mass arises at one loop 
from a diagram with virtual electroweak gauginos and Higgs bosons, giving
\begin{equation}
  m_{\tilde{h}} \,=\, -\frac{\sin 2\beta}{32\pi^2} 
    \left( 3 g^2 M_2 \ln\frac{M_H}{M_2} + g'^2 M_1 \ln\frac{M_H}{M_1} \right),
\label{eq:Higgsinomass}
\end{equation}
where the light Higgs doublet is defined by $h = \sin\beta H_u + \cos\beta 
H_d$, with $\beta$ in the first quadrant, and $M_H$, the mass of $H$, is 
expected to be of order $\tilde{m}$.  (The sign convention of $m_{\tilde{h}}$ 
is such that it agrees with that of $\mu$ in Ref.~\cite{Skands:2003cj} in 
the supersymmetric case.)  This loop-suppressed Higgsino mass, together 
with the environmental selection discussed in the last section, leads to 
the spread spectrum shown in the left panel of Fig.~\ref{fig:spectrum}.

The degeneracy between the charged Higgsino and neutral Higgsinos is lifted 
by electromagnetic corrections and via mixing with the gauginos, to yield 
the mass eigenstates $\chi^+_1$ and $\chi^0_{1,2}$ with masses
\begin{equation}
  m_{\chi^+_1} \,=\, 
    |m_{\tilde{h}}| + \Delta_{\rm EM} + \sin 2\beta\, \frac{M_W^2}{M_2},
\label{eq:chi+mass}
\end{equation}
\begin{equation}
  m_{\chi^0_{1,2}} \,=\, |m_{\tilde{h}}| \mp 
    \left( \frac{1 \pm \sin 2\beta}{2} \right) \xi \, \frac{M_W^2}{M_2},
\label{eq:chi0mass}
\end{equation}
where $\xi \equiv 1 + \tan^2\!\theta_W (M_2/M_1)$ is numerically 
close to unity, and $\theta_W$ is the weak mixing angle. 
$\Delta_{\rm EM} \simeq 310~{\rm MeV}$ represents the electromagnetic 
corrections~\cite{Cirelli:2005uq,Buckley:2009kv}.  The negative sign 
in Eq.~(\ref{eq:Higgsinomass}) is crucial, since it means that the charged 
Higgsino-wino mixing raises the mass of $\chi^+_1$ so that the LSP is 
always neutral.  The numerical size of the shifts of the Higgsino masses 
due to mixing with the gauginos is governed by the Higgsino mass via 
Eq.~(\ref{eq:Higgsinomass})
\begin{equation}
  \frac{M_W^2}{M_2} \,\sim\, 300\, \sin 2\beta 
    \left( \frac{\rm TeV}{|m_{\tilde{h}}|} \right)~{\rm MeV},
\label{mixingsplit}
\end{equation}
and, for $\sin 2\beta \sim 1$ and $|m_{\tilde{h}}| \sim~{\rm TeV}$, 
is the same order as the electromagnetic shift $\Delta_{\rm EM}$.  The 
ordering of the spectrum is $m_{\chi^+_1} > m_{\chi^0_2} > m_{\chi^0_1}$. 
The three gaugino masses and three Higgsino masses depend on just two 
free parameters, $m_{3/2}$ and $\tan\beta$.  (There is also a weak 
logarithmic dependence of the Higgsino masses on $M_H$.)

\begin{figure}[t]
  \center{\includegraphics[scale=1.0]{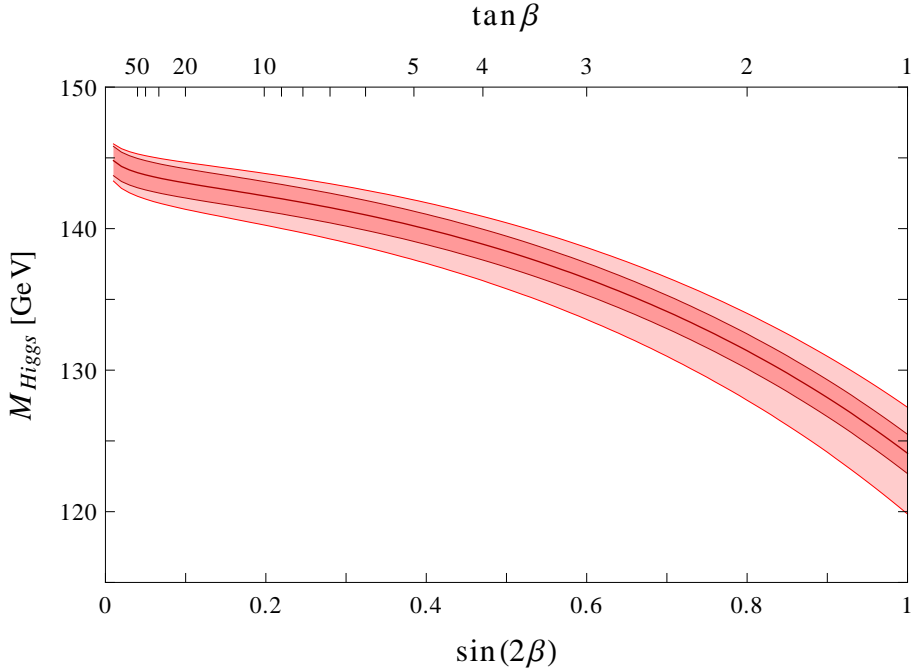}}
\caption{The Higgs mass prediction in Spread Supersymmetry with Higgsino 
 LSP as a function of $\sin 2\beta$. (The corresponding values of 
 $\tan\beta$ are also indicated at the top.)  The solid red curve 
 gives the Higgs mass prediction for $m_t = 173.2~{\rm GeV}$, while 
 the dark-shaded band shows the uncertainty coming from the experimental 
 error of $\delta m_t = \pm 0.9~{\rm GeV}$.  The uncertainty from the 
 superpartner mass scale $\tilde{m}$ is depicted by the light-shaded 
 band, which corresponds to $10^7~{\rm GeV} < \tilde{m} < 10^9~{\rm GeV}$.}
\label{fig:Higgs-mass}
\end{figure}

The present theory allows for a rather robust prediction of the Higgs 
boson mass as a function of angle $\beta$.  This is because the sensitivity 
of the Higgs mass to the superpartner masses is fairly weak, as found 
in Ref.~\cite{Hall:2009nd}, and because the top-squark mixing parameter 
$\theta_{\tilde{t}} \approx A_t/m_{\tilde{t}}$ is very small, of 
order $\epsilon_*/16\pi^2$, as the only source of $A$ terms is anomaly 
mediation, which is suppressed compared with the scalar masses.  In 
Fig.~\ref{fig:Higgs-mass}, we show the Higgs boson mass as a function 
of $\sin 2\beta$, with the masses of the Higgsino, gaugino, and 
scalars taken to be $1~{\rm TeV}$, $10^5~{\rm GeV}$, and $\tilde{m} 
= 10^8~{\rm GeV}$, respectively.
The uncertainty coming from changing $\tilde{m}$ by an order of magnitude 
in both directions is shown as the light-shaded band.  The uncertainty 
from the gaugino masses is of similar size, but generically smaller, 
and the effect of changing the Higgsino mass is negligible.  For the 
top quark mass and QCD coupling, we have used the central values of the 
latest results: $m_t = 173.2 \pm 0.9~{\rm GeV}$~\cite{Tevatron:2011wr} 
and $\alpha_s(M_Z) = 0.1184 \pm 0.0007$~\cite{Bethke:2009jm}.  The 
uncertainty from the error of $m_t$ is depicted by the dark-shaded band, 
with the width corresponding to the $1\sigma$ range.  The uncertainty 
from $\alpha_s$ is smaller.

\begin{figure}[t]
  \center{\includegraphics[scale=1.0]{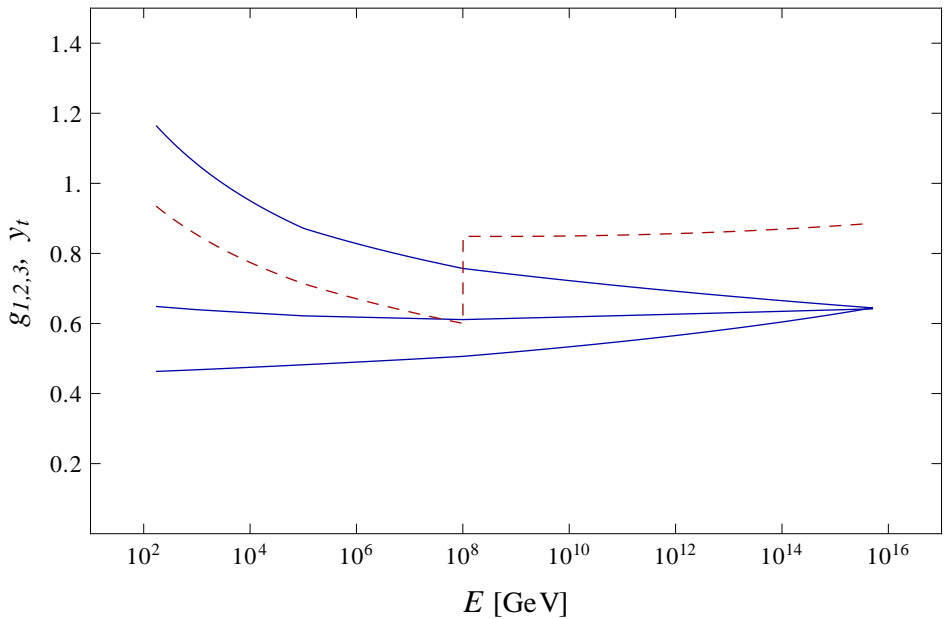}}
\caption{Evolution of the three gauge couplings, $g_{1,2,3}$ (solid, blue) 
 and the top Yukawa coupling (dashed, red) for $\tan\beta = 1$ in Spread 
 Supersymmetry with Higgsino LSP.  The hypercharge gauge coupling has 
 $SU(5)$ normalization.}
\label{fig:couplings}
\end{figure}

Note that the top Yukawa coupling $y_t$ is asymptotically free in the SM, 
so its value at $\tilde{m}$ is smaller than the low energy value.  This 
allows for $\tan\beta = 1$, without encountering a Landau pole below 
the unification scale.  In Fig.~\ref{fig:couplings}, we have plotted 
the running gauge (solid, blue) and top Yukawa (dashed, red) couplings 
as a function of energy for $\tan\beta = 1$.  
The jump in $y_t$ comes from matching the one Higgs (below $\tilde{m}$) 
to two Higgs (above $\tilde{m}$) theories.  The top Yukawa coupling 
is well perturbative at the unification scale.  Requiring that all the 
Yukawa couplings are perturbative up to the unification scale, we find 
$0.7 \simlt \tan\beta \simlt 100$.  Therefore, the range of $\sin 2\beta$ 
consistent with perturbative gauge coupling unification is
\begin{equation}
  0.02 \,\simlt\, \sin 2\beta \,\leq\, 1,
\label{eq:sin-2beta}
\end{equation}
which spans almost the entire range of Fig.~\ref{fig:Higgs-mass}.

\subsection{Unification}
\label{subsec:unification}

As can be seen in Fig.~\ref{fig:couplings}, unification of the three 
SM gauge couplings works very well in Spread Supersymmetry.  To quantify 
it, let us consider the size of the threshold correction $\delta(E)$ 
required for gauge coupling unification at energy $E$, where 
$\delta \equiv \sqrt{(g_1^2-\bar{g}^2)^2 + (g_2^2-\bar{g}^2)^2 
+ (g_3^2-\bar{g}^2)^2}/\bar{g}^2$ with $\bar{g}^2 \equiv 
(g_1^2+g_2^2+g_3^2)/3$.  In Spread Supersymmetry with Higgsino LSP, 
this quantity takes a minimum value at
\begin{equation}
  M_{\rm unif} \,\simeq\, (5~\mbox{--}~8) \times 10^{15}~{\rm GeV}, 
\label{eq:M-unif}
\end{equation}
with
\begin{equation}
  \delta(M_{\rm unif}) \,\simeq\, 0.004~\mbox{--}~0.008,
\label{eq:delta-Munif}
\end{equation}
where the value of $M_{\rm unif}$ is most sensitive to the gaugino masses 
(anti-correlation), while that of $\delta(M_{\rm unif})$ to the Higgsino 
mass (correlation).  The result of Eq.~(\ref{eq:delta-Munif}) can be 
compared with the values in the SM and in the MSSM:\ $\delta_{\rm min,\,SM} 
\simeq 0.06$ and $\delta_{\rm min,\,MSSM} \simeq O(0.01)$ (depending on 
the superpartner spectrum).  Spread Supersymmetry, therefore, achieves 
gauge coupling unification, at least, at a level of the MSSM.

The value of the unified gauge coupling at $E \simeq M_{\rm unif}$ is
\begin{equation}
  g_{\rm unif}(M_{\rm unif}) \,\simeq\, 0.65.
\label{eq:g-unif}
\end{equation}
This and Eq.~(\ref{eq:M-unif}) have an important implication on the 
rate of dimension six proton decay, caused by an exchange of the 
unified gauge bosons.  Since the (partial) decay rate is proportional 
to $g_{\rm unif}(M_{\rm unif})^4/M_{\rm unif}^4$, we find
\begin{equation}
  \frac{\Gamma_{\rm Spread}}{\Gamma_{\rm MSSM}} 
    \,\simeq\, 30~\mbox{--}~200,
\label{eq:Gamma-ratio}
\end{equation}
where we have used $g_{\rm unif}(M_{\rm unif})|_{\rm MSSM} \simeq 0.7$ 
and $M_{\rm unif}|_{\rm MSSM} \simeq 2 \times 10^{16}~{\rm GeV}$.  This 
corresponds to the lifetime~\cite{Hisano:2000dg}
\begin{equation}
  \tau_{p \rightarrow e^+\pi^0} \,\simeq\, 
    (0.8~\mbox{--}~5) \times 10^{34}~{\rm years}
\label{eq:p-e-pi0}
\end{equation}
in 4-dimensional supersymmetric grand unified theories.  This 
range is just above the current lower limit from Super-Kamiokande 
$\tau_{p \rightarrow e^+\pi^0} > 8.2 \times \times 10^{33}~{\rm 
years}$~\cite{Super-K:2009gd}, and can be fully covered by 
the planned Hyper-Kamiokande experiment at the $3\sigma$ 
level~\cite{Abe:2011ts}.  If grand unification is realized 
in higher dimensions~\cite{Kawamura:2000ev}, dimension six proton 
decay can have a variety of final states~\cite{Nomura:2001tn}. 
The lifetime in this case is also expected to be shorter than the 
corresponding case in which the low energy theory is the MSSM.

\subsection{Dark matter}
\label{subsec:dark matter}

If dark matter is composed only of Higgsinos then the freeze-out mechanism 
requires the Higgsino mass to be $1.1~{\rm TeV}$.  On the other hand, 
with multi-component dark matter the Higgsino fraction, and therefore 
the Higgsino mass, depends on the relevant multiverse distribution 
functions.  This makes the mass of the Higgsino lighter:
\begin{equation}
  m_{\tilde{h}} \,\simeq\, 1.1 
    \left(\frac{\Omega_{\tilde{h}}}{\Omega_{\rm DM}} \right)^{1/2}\!{\rm TeV}.
\label{eq:Higgsino-mass}
\end{equation}
(In this subsection, we ignore the small difference between $m_{\tilde{h}}$ 
and the LSP mass $m_{\chi^0_1}$.)

For illustration, let us take a distribution function $f(\tilde{m}) 
d\tilde{m} = \tilde{m}^p (d\tilde{m}/\tilde{m})$ for values of $\tilde{m}$ 
giving a Higgsino mass in the region corresponding to the critical 
environmental boundary of Eq.~(\ref{eq:xi-bound}).  This includes 
a quadratic weighting factor resulting from the environmental requirement 
that the Higgs vacuum expectation value is below its critical value. 
For mixed Higgsino/axion dark matter, the environmental boundary of 
Eq.~(\ref{eq:xi-bound}) can be cast in the form
\begin{equation}
  x^2 + y^2 \,<\, 1;
\qquad\quad
  x \,=\, \frac{\tilde{m}}{\tilde{m}_{\rm c}} 
  \,=\, \frac{m_{\tilde{h}}}{m_{\tilde{h}{\rm c}}},
\qquad\quad
  y \,=\, \frac{\theta}{\theta_{\rm c}},
\label{Higgsino-axion boundary}
\end{equation}
where $\theta$ is the axion misalignment angle, whose multiverse 
distribution is expected to be flat, and subscripts refer to the 
critical values.  As an example, $p=1$ leads to equal multiverse 
averages of Higgsino and axion dark matter:\ $\langle \Omega_{\tilde{h}} 
\rangle = \langle \Omega_a \rangle$, giving an average Higgsino mass 
of about $700~{\rm GeV}$.  Lower values of $p$ give lighter Higgsinos; 
if $p = \epsilon \ll 1$, Higgsinos account on average for only a fraction 
$\epsilon$ of the dark matter, giving an expected value of the Higgsino 
mass of $\approx \sqrt{\epsilon}~{\rm TeV}$.  Ultimately, experiments 
will see a correlation between the Higgsino mass and the fractions 
of Higgsino and axion dark matter.

In Spread Supersymmetry with a Higgsino-like LSP, the direct detection 
of dark matter is challenging.  The mass splitting between the 
various components of the Higgsino, Eqs.~(\ref{eq:chi+mass}) and 
(\ref{eq:chi0mass}), are sufficient that the LSP-nucleus scattering 
occurs as the elastic scatter of a Majorana fermion.  In the limit 
that the gaugino component of $\chi^0_1$ is ignored, the spin-independent 
cross section for scattering from nuclei has been studied in 
Refs.~\cite{Cirelli:2005uq,Essig:2007az,Hisano:2011cs}.  In 
Ref.~\cite{Cirelli:2005uq}, 1-loop electroweak gauge boson 
contributions (from both box and Higgs exchange diagrams) led to 
a scattering cross section from protons of $\sigma_p \sim 2 \times 
10^{-46}~{\rm cm}^2$ for a Higgsino mass of $1~{\rm TeV}$ and a Higgs 
mass of $115~{\rm GeV}$.  In Ref.~\cite{Essig:2007az}, diagrams involving 
gluons were included leading to a similar result for $\sigma_p$. 
However, the authors of Ref.~\cite{Hisano:2011cs} find an additional 
twist~2 operator contribution, and disagree with certain results of 
the earlier papers, concluding that for Higgs masses in the predicted 
range of Fig.~\ref{fig:Higgs-mass} there is a cancellation between the 
Higgs exchange and non-Higgs exchange diagrams, leading to $\sigma_p 
\sim 10^{-48}~{\rm cm}^2$.

Our LSP, however, contains a gaugino component, so additional contributions 
to the spin-independent scattering from nuclei arise from a tree-level 
Higgs exchange amplitude proportional to the small wino component. 
This contribution depends on $\tan\beta$ and has been studied in 
Ref.~\cite{Hisano:2004pv}.  For gaugino masses of interest to us, this 
contribution to $\sigma_p$ is of order $10^{-47}~{\rm cm}^2$ or smaller. 
Further study should investigate the effect of uncertainties of the 
nuclear matrix elements on cancellations between the varying amplitudes. 
The spin-dependent cross section is generically dominated at tree 
level, and is in the range of $O(10^{-46}~\mbox{--}~10^{-44}~{\rm 
cm}^2)$~\cite{Hisano:2004pv}.

The indirect detection of Higgsino dark matter may occur via the detection 
of monochromatic photons of energy $m_{\tilde{h}}/2$ arising from 
$\tilde{h}\tilde{h} \rightarrow \gamma\gamma$ in the halo of our galaxy. 
For $m_{\tilde{h}} = 100~{\rm GeV}~\mbox{--}~1~{\rm TeV}$, the one-loop 
electroweak annihilation cross section is $\sigma v (\tilde{h}\tilde{h} 
\rightarrow \gamma\gamma) \approx 10^{-28}~{\rm cm^3/s}$ and has 
no significant Sommerfeld enhancement.  Using the NFW profile 
for the distribution of galactic dark matter, data from the Fermi~LAT 
place an upper limit on this cross section of $(20 - 100) \times 
10^{-28}~{\rm cm^3/s}$ for $m_{\tilde{h}}$ in the range of 
$(100~\mbox{--}~300)~{\rm GeV}$~\cite{Abdo:2010nc}.  The eventual 
discovery of a monochromatic galactic photon signal would directly 
yield the Higgsino mass, motivating the construction of a linear 
lepton collider with a definite energy.  In addition, from this mass 
one could accurately infer the fraction of dark matter in LSPs.

\subsection{Collider signals of Higgsinos}
\label{subsec:collider}

Pairs of $(\chi^+_1, \chi^0_1, \chi^0_2)$ will be produced by the 
Drell-Yan mechanism at LHC.  The charged state beta decays to the neutral 
states, but the mass difference is greater than $\approx 300~{\rm MeV}$ 
allowing the two-body decay $\chi^+_1 \rightarrow \chi^0_{1,2} \pi^+$. 
This gives $c\tau \simlt 1~{\rm cm}$, so that the charged tracks of 
$\chi^+_1$ are not visible.  For $|m_{\tilde{h}}| \sim {\rm TeV}$, 
the pion is too soft to see above backgrounds.  Similarly, the decays 
$\chi^0_2 \rightarrow \chi^0_1 e^+ e^-$ give $e^+ e^-$ pairs that are 
too soft to see above backgrounds.  However, for lower $|m_{\tilde{h}}|$, 
the mass splittings between the states may be large for $\sin 2\beta \sim 
1$ (see Eq.~(\ref{mixingsplit})), so that events with decays of boosted 
$\chi^+_1$ or $\chi^0_2$ states could give observable LHC signals.

At a future lepton collider, signals of nearly degenerate Higgsinos result 
from $\chi^+_1 \chi^-_1 \gamma$ and $\chi^0_1 \chi^0_2 \gamma$ productions 
followed by $\chi^\pm_1 \rightarrow \chi^0_{1,2} \pi^\pm$ and $\chi^0_2 
\rightarrow \chi^0_1 e^+ e^-$ decays~\cite{Chattopadhyay:2005mv}.  Hard 
initial state photon radiation is required since otherwise the event 
contains only the soft products of $\chi^+_1$ and $\chi^0_2$ decays 
and such events are overwhelmed by underlying events from the collision 
of beamstrahlung photons that produce soft particles.  The SM background 
arising from $\nu \bar{\nu} \gamma$ production has a cross section about 
three orders of magnitude larger than that of $\chi^+_1 \chi^-_1 \gamma$ 
and $\chi^0_1 \chi^0_2 \gamma$ productions for a Higgsino mass in the 
region of $1~{\rm TeV}$.  Although this background can be reduced by 
using polarized beams, search strategies should be devised that involve 
the soft $\pi^+$ and $e^+ e^-$ from $\chi^+_1$ and $\chi^0_2$ decays.
Although $\chi^+_1$ has $c \tau \sim$ cm, it may be possible to observe 
the charged tracks if the boost factor $\sqrt{s}/2m_{\tilde{h}}$ is 
sufficiently large and if the luminosity is large enough to yield events 
with decays occurring at times of a few $\tau$.

Precision measurements of the overall Higgsino mass and the two mass 
splittings, Eqs.~(\ref{eq:chi+mass}) and (\ref{eq:chi0mass}), would 
be a powerful probe of the theory.  These three observables depend on 
only $m_{3/2}$, $\sin 2\beta$ and $\ln M_H/M_2$, which would be fit to 
the data.   Furthermore, $\sin 2\beta$ is correlated with the Higgs mass 
prediction, Fig.~\ref{fig:Higgs-mass}, yielding a consistency check. 
One would then infer the gaugino spectrum.  Observing splittings amongst 
the Higgsinos of a few hundred MeV would imply electroweak gaugino 
masses of order $100~{\rm TeV}$ independent of the underlying theory. 
Such a large breaking of supersymmetry would imply a very large, but 
model dependent, fine-tuning in electroweak symmetry breaking.  Within 
Spread Supersymmetry the amount of fine-tuning depends on the value of 
$\epsilon_*$ inferred from $m_{3/2}$ and $\ln M_H/M_2$.  For $\epsilon_* 
\sim 10^{-2}$ the tuning is of order $1$ part in $\approx 10^{14}$.

\section{Spread Supersymmetry with Wino LSP}
\label{sec:WinoLSP}

In this section we consider the case where supergravity effects yield 
a Higgsino mass of order $m_{3/2}$.  This can arise from a Higgs 
bilinear term in the K\"{a}hler potential $K = \lambda H_u H_d 
+ {\rm h.c.}$~\cite{Giudice:1988yz} or by a vacuum readjustment 
induced by supergravity corrections to the potential of flat 
supersymmetry~\cite{Hall:1983iz,Hempfling:1994ae}.  The superparticle 
spectrum becomes that in the right panel of Fig.~\ref{fig:spectrum}:
\begin{equation}
  m_{\tilde{q},\tilde{l},H^{0,\pm},A} \,\sim\, \tilde{m},
\qquad
  m_{\tilde{h},\tilde{G}} \,\sim\, \epsilon_* \tilde{m},
\qquad
  m_{\tilde{g},\tilde{W},\tilde{B}} \,\sim\, 
    \frac{\epsilon_*}{16\pi^2} \tilde{m},
\label{eq:spectrum-2}
\end{equation}
where the gaugino masses are generated by anomaly mediation as well as 
one loop of the Higgs-Higgsino.  This, therefore, leads to a scenario 
similar to the one discussed in Refs.~\cite{Giudice:1998xp,Wells:2003tf}.%
\footnote{These papers did not discriminate between the masses of 
 scalars and the gravitino, $m_{\tilde{q},\tilde{l},H^{0,\pm},A} \sim 
 m_{\tilde{G}}$, corresponding to the case $\epsilon_* \approx O(1)$ 
 in our scenario.}

We assume that the wino is the LSP, which is the case unless the 
contribution from a Higgs-Higgsino loop dominates over that from anomaly 
mediation.  Phenomenology of this theory is mostly as discussed in 
Ref.~\cite{Wells:2003tf}.  The Higgs boson mass depends on the gaugino 
and Higgsino masses as well as $\tan\beta$ and $\epsilon_*$, and 
is generically in the range
\begin{equation}
  M_{\rm Higgs} \,\approx\, (110~\mbox{--}~140)~{\rm GeV}.
\label{eq:Higgs-theory2}
\end{equation}
Gauge coupling unification works essentially as in the MSSM; we find
\begin{equation}
  \delta(M_{\rm unif}) \,\simlt\, O(0.01),
\qquad
  M_{\rm unif} \,\simeq\, 10^{16}~{\rm GeV}.
\label{eq:gcu-theory2}
\end{equation}
The lightest gaugino is typically the wino.  The mass splitting between 
the charged and neutral components, $\varDelta m \equiv m_{\chi^+} 
- m_{\chi^0}$, is given by
\begin{equation}
  \varDelta m \,=\, \Delta_{\rm EM} 
      + O\biggl(\frac{m_W^2 m_{\tilde{W}}}{m_{\tilde{h}}^2}\biggr) 
    \,\simeq\, \Delta_{\rm EM},
\label{eq:splitting-theory2}
\end{equation}
where $\Delta_{\rm EM} \simeq 160~{\rm MeV}$ is the electromagnetic 
contribution.  The second contribution to $\varDelta m$ in 
Eq.~(\ref{eq:splitting-theory2}) from mixing with the Higgsino 
is expected to be less than $1~{\rm MeV}$ in our framework.  If 
the wino LSP composes the entire dark matter, then $m_{\tilde{W}} 
\simeq (2.7~\mbox{--}~3.0)~{\rm TeV}$~\cite{Hisano:2006nn}.  If 
the axion composes a significant part of dark matter, the mass 
of the wino becomes correspondingly smaller:
\begin{equation}
  m_{\tilde{W}} \,\simeq\, (2.7~\mbox{--}~3.0)
    \left(\frac{\Omega_{\tilde{W}}}{\Omega_{\rm DM}} \right)^{1/2}\!{\rm TeV},
\label{eq:wino-mass}
\end{equation}
ignoring the variation of the Sommerfeld enhancement factor with the LSP 
mass.  The fraction $\Omega_{\tilde{W}}/\Omega_{\rm DM}$ is determined 
by the multiverse distribution function $f(\tilde{m})$.

The cross section for the direct detection of wino dark matter is 
generically an order of magnitude larger than for Higgsino dark 
matter.  In the limit of neglecting the Higgsino component of the 
LSP, and ignoring contributions from operators involving gluons, 
Ref.~\cite{Cirelli:2005uq} quotes a spin-independent cross section from 
the proton of $\sigma_p \sim 10^{-45}~{\rm cm}^2$ for a wino mass of 
$2.4~{\rm TeV}$ and a Higgs mass of $115~{\rm GeV}$.  However, including 
the gluon operators the authors of Refs.~\cite{Hisano:2010ct,Hisano:2011cs} 
find a cancellation between the Higgs exchange and non-Higgs exchange 
diagrams, leading to $\sigma_p \sim O(10^{-47}~{\rm cm}^2)$ for a 
Higgs mass in the range of $\approx (114~\mbox{--}~140)~{\rm GeV}$ 
and $m_{\tilde{W}} \sim 3~{\rm TeV}$.  The tree-level Higgs exchange 
contribution, involving mixing with the Higgsino, is very small, $\simlt 
10^{-48}~{\rm cm}^2$, in the parameter region of interest.  The spin-dependent 
cross section is dominated by loop diagrams and is in the range of 
$O(10^{-46}~\mbox{--}~10^{-44})~{\rm cm}^2$~\cite{Hisano:2011cs}.

The indirect detection of wino dark matter from annihilation to photons 
is promising.  The cross section $\sigma v(\tilde{W}\tilde{W} \rightarrow 
\gamma\gamma)$ at one loop is about $10^{-27}~{\rm cm^3/s}$, about 
an order of magnitude larger than for the case of pure Higgsino dark 
matter.  Furthermore, while the Sommerfeld enhancement occurs at too 
large a mass to be relevant for Higgsinos, in the wino case the resonance 
occurs at a wino mass of about $2.3~{\rm TeV}$ and yields an enhancement 
of the cross section by factors of $(3,\,30,\,125)$ for wino masses of 
$(1,\,2,\,2.5)~{\rm TeV}$, respectively~\cite{Hisano:2004ds}.  In both 
the Higgsino and wino LSP cases, a measurement of the energy of the 
photon line allows an inference of the fraction of dark matter carried 
by the LSP and motivates the construction of a lepton collider.

Prospects for detecting the wino LSP at the LHC are better than that 
for the Higgsino LSP.  This is mainly because the mass splitting 
between the charged and neutral components is smaller, $\varDelta m 
\simeq 160~{\rm MeV}$, so that the decay length of $\chi^+ \rightarrow 
\chi^0 \pi^+$ is longer, $c\tau \approx O(10~{\rm cm})$.  Drell-Yan 
production of $\chi^+ \chi^-$ yields events with (disappearing) charged 
tracks, that can be triggered by high $p_T$ jets or missing transverse 
energy~\cite{Ibe:2006de,Buckley:2009kv}.  For $\sqrt{s} = 13~{\rm TeV}$ 
and integrated luminosity of $10~{\rm fb^{-1}}$ ($100~{\rm fb^{-1}}$), 
reach for the wino mass is estimated to be $\approx 350~{\rm GeV}$ 
($550~{\rm GeV}$)~\cite{Buckley:2009kv}.

The wino could be discovered at a future lepton collider via very 
distinctive events arising from the production of $\chi^+ \chi^-$ followed 
by the decay $\chi^\pm \rightarrow \chi^0 \pi^\pm$~\cite{Gunion:2001fu}. 
The highly ionizing $\chi^\pm$ tracks are crucial, since soft $\pi^+ \pi^-$ 
events have a large background from two photon induced processes.  With 
$c\tau \sim 10~{\rm cm}$, independent of the wino mass, these ionizing 
tracks will be seen to end.  Furthermore, a very soft pion will intersect 
the end of the track at large angle, and such events have large missing 
energy.  Measuring the wino mass would allow the fraction of dark matter 
in winos to be inferred.  Measuring the $\chi^+$ lifetime would lead 
to a determination of $\varDelta m$ limiting the size of the Higgsino 
mixing contribution to the wino mass.  For example, an observation that 
$\varDelta m < 170~{\rm MeV}$ would imply that the Higgsino is heavier 
than the wino by at least a factor of $30 \sqrt{m_{\tilde{W}}/{\rm TeV}}$ 
and hence that there is a very high degree of fine-tuning 
in electroweak symmetry breaking, greater than one part in 
$10^5\,(m_{\tilde{W}}/{\rm TeV})^3$.

\section{The Environmental MSSM}
\label{sec:E-MSSM}

Suppose the supersymmetry breaking field $X$ is neutral, rather than 
charged, so that all SM superpartners (including the Higgsino) acquire 
masses directly from the hidden sector at order $\tilde{m}$, through 
operators suppressed by $M_*$.  In general, there can be mild hierarchies 
among these superpartners, so that there will be a great deal of model 
dependence in the precise superpartner spectrum and the nature of the 
LOSP.  It is, however, still true that there is generically a large 
forbidden window of $\tilde{m}$, as illustrated in the right panel of 
Fig.~\ref{fig:tilde-m}.  

The key point is that the normalization of the spectrum will arise from 
the environmental bound resulting from the forbidden shaded zone of 
Fig.~\ref{fig:tilde-m}.  The gravitino mass is $m_{3/2} = F_X/\sqrt{3} 
M_{\rm Pl} \equiv \epsilon_* \tilde{m}$, and we expect the gravitino 
to be the LSP for all but the highest values of the messenger scale, 
$M_* \simgt \sqrt{3} M_{\rm Pl}$.  As $\epsilon_*$ is reduced, the 
gravitino mass gets smaller, implying that the environmental bound 
on the LOSP freeze-out abundance gets milder and the entire SM superpartners 
spectrum gets heavier.  We do not consider this limit, since the bound 
on the reheating temperature from thermal production of gravitinos 
becomes more powerful $T_R < 10^9~{\rm GeV} (m_{3/2}/10~{\rm GeV})$, 
narrowing the range of the forbidden window.  For $\epsilon_* \sim 
10^{-2}$ the SM superpartner masses are all expected to be 1 to 2 
orders of magnitude above the weak scale.  Hence the lightest 
Higgs boson mass is predicted to be roughly in the range $\approx 
100~\mbox{--}~130~{\rm GeV}$.

As one example of parameters for the Environmental MSSM, consider 
$\tilde{m} \sim 10~{\rm TeV}$, a LOSP mass somewhat larger than a TeV 
and $\epsilon_* \sim 10^{-3}$, giving $m_{3/2} \sim 10~{\rm GeV}$. 
For LOSP masses lighter than about a TeV, gravitinos arising from 
LOSP freeze-out and decay cannot comprise all the dark matter, because 
the LOSP decay upsets Big Bang Nucleosynthesis (BBN)~\cite{Kawasaki:2008qe}. 
In our scheme, however, the LOSP mass may be lighter than a TeV without 
being excluded by BBN because the gravitino may only be a sub-dominant 
component of dark matter.  For LOSP masses above a TeV, dark matter 
could dominantly be gravitinos arising from LOSP freeze-out and 
decay, since the LOSP decays more rapidly and evades the BBN limits. 
Such decays of a $\tilde{\tau}$ LOSP could solve the BBN lithium 
problem~\cite{Bailly:2008yy}.

If $\epsilon_* \simgt 1$, the LSP can be the LOSP---the superpartner 
spectrum is then as in the standard MSSM with gravity mediation (defined 
broadly)~\cite{Elor:2009jp}.  The scale of the superparticles, however, 
is now determined not by fine-tuning of electroweak symmetry breaking, 
but by the LSP relic abundance: $\Omega_{\rm LSP} < \Omega_{\rm DM}$. 
If the LSP contains a significant amount of the Higgsino or wino, this 
can lead to relatively heavy superparticles, which however may still 
be within reach of the LHC.

\section{Scanning of {\boldmath $\mu$}}
\label{sec:mu-scanning}

So far, we have assumed that the supersymmetric Higgsino mass parameter, 
$\mu$, is absent from the superpotential.  In this section we consider 
scanning $\mu$ and $\tilde{m}$ independently.

If the multiverse distribution is logarithmic in $\mu$ (or favors small 
values), then typical observers live in universes with $\mu \ll \tilde{m}$ 
because electroweak symmetry does not break for $\mu \simgt \tilde{m}$ 
(so that no/few observers arise).  In this case the Higgsino mass will 
be dominated by other sources.  An effective $\mu$ term arises either 
from $K = \lambda H_u H_d$ or vacuum readjustment, leading to the theory 
discussed in Section~\ref{sec:WinoLSP}, or is generated radiatively, 
leading to the theory described in section \ref{sec:HiggsinoLSP}.

What if the a priori multiverse distribution for $\mu$ favors large $\mu$? 
In this case typical observers will see $\mu$ of order $\tilde{m}$, close 
to the upper limit imposed by the requirement of electroweak symmetry 
breaking.  This is a multiverse solution to the $\mu$ problem and, in 
the context of Spread Supersymmetry (i.e.\ if $X$ is charged), yields 
a wino LSP with a supersymmetric spectrum as illustrated in the right 
panel of Fig.~\ref{fig:spectrum}, except that the Higgsinos now have 
a mass comparable to the scalar superpartners rather than to the 
gravitino.  The collider and dark matter phenomenology is as in 
section~\ref{sec:WinoLSP}.  If $X$ is neutral, the resulting theory 
is the Environmental MSSM discussed in the previous section.  In either 
theory, the distribution for $\mu$ now contributes to the effective 
multiverse distribution that favors large values for supersymmetry 
breaking.  The fine-tuning in the Higgs mass parameter can be overcome 
by a combination of the $\mu$ and $\tilde{m}$ distributions, making 
it more plausible that $\tilde{m}$ is typically (much) larger than 
the weak scale.

\section{Conclusions}
\label{sec:concl}

If a Standard Model Higgs is discovered at the LHC in the region of 
$115~\mbox{--}~145~{\rm GeV}$, with no signs of any other new physics, 
the possibility that the weak scale is determined by environmental 
selection on a multiverse will be increased.  Does this mean that 
no new physics is expected in the TeV domain?

An environmental upper bound on the amount of dark matter in the universe 
can be phrased as an upper limit on the temperature of matter-radiation 
equality
\begin{equation}
  T_{\rm eq} \, < \, T_{\rm eq,c},
 \label{eq:Te}
\end{equation}
with the critical value, $T_{\rm eq,c}$, not far above the observed value. 
In this case, any freeze-out relic of mass $m$, with annihilation cross 
section within a few orders of magnitude of $m^{-2}$, will satisfy 
$m < m_{\rm c} \approx {\rm TeV}$.   Furthermore, if the multiverse 
distribution, including selection effects from other boundaries, favors 
large values of $m$, typical observers will find the mass of this weakly 
interacting massive particle (WIMP) close to its critical value.  This 
quite generally leads to the possibility of new physics associated with 
WIMPs at the TeV scale, without having a direct connection to electroweak 
symmetry breaking.  A key difference from conventional WIMPs is that 
the environmental boundary of Eq.~(\ref{eq:Te}) limits all cold relics 
suggesting multi-component dark matter, for example axions and WIMPs.

Specializing to supersymmetric theories, if the overall scale of 
supersymmetry breaking scans, with fixed superpartner mass ratios, 
the normalization of the spectrum may by determined by Eq.~(\ref{eq:Te}), 
fixing the LSP mass to be of order $1~{\rm TeV}$.  This could lead to 
``Environmental Supersymmetry'' with a spectrum of superpartners in 
the multi-TeV domain, $1~\mbox{--}~2$ orders of magnitude larger than 
expected from natural theories of weak-scale supersymmetry.  Alternatively, 
if the leading supersymmetry breaking arises at quadratic order in 
the $F_X$ spurion, a hierarchical spectrum of superpartners results, 
giving ``Spread Supersymmetry.''  We have studied the spectra, dark 
matter and collider signals of two examples of Spread Supersymmetry 
having Higgsino and wino LSPs with scalar superpartner masses of 
order $10^6$ and $10^4~{\rm TeV}$, respectively.  The examples studied 
have the same set of fields as the MSSM, but this is not necessary; 
a similar pattern of spectra would result in extended theories, 
for example with singlets added.

Signals for such theories may show up first via indirect detection of 
galactic dark matter through a monochromatic photon signal, for example 
for a pure Higgsino or wino LSP, or via direct detection of galactic 
dark matter, for example for the case of a mixed singlino/Higgsino LSP. 
The former case would motivate the construction of a lepton collider 
with energy optimized for the particular energy of the photon signal, 
while the latter would motivate very high luminosity LHC studies. 
A key goal would be to learn sufficient about the structure of the 
underlying theory to demonstrate that electroweak symmetry breaking 
has a very high degree of fine-tuning, pointing to a multiverse.

The simplest and most natural theories of weak-scale supersymmetry 
lead to a Higgs boson lighter than about $100~{\rm GeV}$; of course, 
theories can be extended and naturalness can be relaxed to accommodate 
a heavier Higgs.  On the contrary, the theories introduced in this 
paper {\it predict} a heavier Higgs in the mass range $\approx 
120~\mbox{--}~145$, $110~\mbox{--}~140$, and $100~\mbox{--}~130~{\rm GeV}$ 
for Higgsino LSP Spread Supersymmetry, Wino LSP Spread Supersymmetry, 
and the Environmental MSSM, respectively.   We expect this Higgs boson 
to be indistinguishable from the Standard Model Higgs, and for colored 
superpartners to be out of reach for the LHC.

\section*{Acknowledgments}

This work was supported in part by the Director, Office of Science, Office 
of High Energy and Nuclear Physics, of the US Department of Energy under 
Contract DE-AC02-05CH11231 and by the National Science Foundation under 
grants PHY-1002399 and PHY-0855653.

\end{document}